\documentclass[times,10pt,onecolumn]{article}

\usepackage{epsf}\epsfverbosetrue
\usepackage{graphics,epsfig,color}
\usepackage{graphicx,epsfig} 
\usepackage{multirow}
\usepackage{alltt}
\usepackage{subfigure}
\usepackage{float}
\usepackage{url}
\usepackage{graphicx}
\usepackage{verbatim} 
\usepackage{algorithm}
\usepackage{algpseudocode}
\usepackage{footnote} 
\usepackage{latexsym} 
\usepackage{amsmath}
\usepackage{sidecap} 
\usepackage{color}

\begin{document}

\title{A Tutorial on the Implementation of Ad-hoc On Demand Distance Vector (AODV) Protocol in Network Simulator (NS-2)\vspace{50pt}}

\author{Mubashir Husain Rehmani, \\ Sidney Doria, and Mustapha Reda Senouci \vspace{170pt} \thanks{M. H. Rehmani is with INRIA, France, e-mail: mubashir.rehmani@inria.fr; S. Doria is with UFCG, Brazil, e-mail: sidney@dsc.ufcg.edu.br; M. R. Senouci is with Laboratory of Research in Artificial Intelligence, Algeria, email: mrsenouci@gmail.com; Special thanks to Hajer Ferjani, she has a Masters Degree in Networking from the National School of Computer Science of Tunisia, CRISTAL Laboratory in 2006, Tunisia, e-mail: 
f.hajer@gmail.com. The author would like to thanks Aline Carneiro Viana, who is with INRIA, France, e-mail: aline.viana@inria.fr;} }

\date{Version 1 \vspace{55pt} \\ $28^{th}$ June 2009}

\maketitle
\pagebreak
\thispagestyle{plain}

\tableofcontents
\pagebreak

\begin{abstract}
The Network Simulator (NS-2) is a most widely used network simulator. It has the capabilities to simulate a range of networks including wired and wireless networks. In this tutorial, we present the implementation of Ad Hoc On-Demand Distance Vector (AODV) Protocol in NS-2. This tutorial is targeted to the novice user who wants to understand the implementation of AODV Protocol in NS-2.
\end{abstract}


\section{Introduction}
\label{sec:introduction}

The Network Simulator (NS-2) \cite{IEEEhowto:ns} is a most widely used network simulator. This tutorial presents the implementation of Ad Hoc On-Demand Distance Vector (AODV) Protocol \cite{IEEEhowto:aodv} in NS-2. The expected audience are students who want to understand the code of AODV and researchers who want to extend the AODV protocol or create new routing protocols in NS-2. The version considered is NS-2.32 and 2.33, but it might be useful to other versions as well. Throughout the rest of this tutorial, the under considered files are aodv.cc, aodv.h, aodv\_logs.cc, aodv\_packet.h, aodv\_rqueue.cc, aodv\_rqueue.h, aodv\_rtable.cc, aodv\_rtable.h which can be found in AODV folder in the NS-2 base directory.


\section{File Dependency of AODV Protocol}
\label{dep}
\begin{figure}[htbp]
    \begin{center}
    \includegraphics[width=9cm]{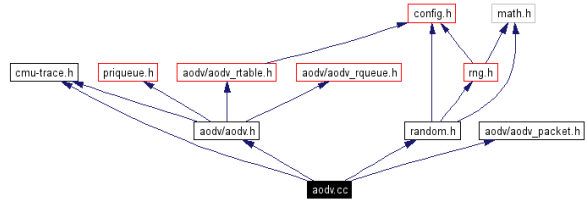}
\vspace{-0.4cm} \caption{File Reference of `AODV.CC'. \vspace{-1.5cm}} 
    \label{fig0}
\end{center}
\end{figure}

\begin{figure}[htbp]
    \begin{center}
    \includegraphics[width=9cm]{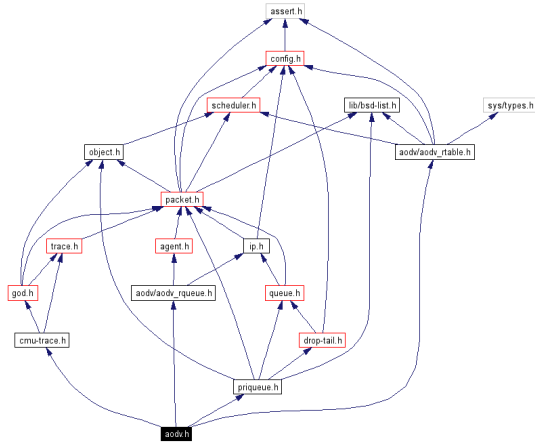}
\vspace{-0.4cm} \caption{File Reference of `AODV.H'.} \vspace{-0.7cm} 
    \label{fig1}
\end{center}
\end{figure}

Fig.~\ref{fig0} and~\ref{fig1} shows the file dependency of AODV Protocol~\cite{ref}. As AODV is a routing protocol, so it is derived from the class {\it Agent}, see agent.h.

\section{Flow of AODV}
\label{sec:flow}

In this section, we describes the general flow of AODV protocol through a simple example:

\begin{enumerate}

\item In the TCL script, when the user configures AODV as a routing protocol by using the command,\\
\$ns node-config -adhocRouting AODV

the pointer moves to the ``start'' and this ``start'' moves the pointer to the Command function of AODV protocol.
\item In the Command function, the user can find two timers in the ``start'' \\
    * btimer.handle((Event*) 0);\\
    * htimer.handle((Event*) 0);
\item Let's consider the case of htimer, the flow points to HelloTimer::handle(Event*) function and the user can see the following lines:
   \begin{quote}
 agent -$ >$ sendHello();\\
   double interval = MinHelloInterval +
                 ((MaxHelloInterval - MinHelloInterval) *
Random::uniform());\\
   assert(interval -$>$ = 0);\\
   Scheduler::instance().schedule(this, $\&$intr, interval);
\end{quote}
These lines are calling the sendHello() function by setting the appropriate interval of Hello Packets.
\item Now, the pointer is in AODV::sendHello() function and the user can see Scheduler::instance().schedule(target\_, p, 0.0) which will schedule the packets.
\item In the destination node AODV::recv(Packet*p, Handler*) is called, but actually this is done after the node is receiving a packet.
\item AODV::recv(Packet*p, Handler*) function then calls the recvAODV(p) function.
\item Hence, the flow goes to the AODV::recvAODV(Packet *p) function, which will check different packets types and call the respective function.
\item In this example, flow can go to\\
case AODVTYPE\_HELLO: \\
   recvHello(p);\\
   break;
\item Finally, in the recvHello() function, the packet is received.

\end{enumerate}

\section{Trace Format of AODV}
\label{sec:trace}

In NS-2, the general trace format is given as below:\\

s 0.000000000 \_0\_ RTR  --- 0 AODV 44 [0 0 0 0] ------- [0:255 -1:255 1 0] [0x1 1 [0 2] 4.000000] (HELLO)\\

s 10.000000000 \_0\_ RTR  --- 0 AODV 48 [0 0 0 0] ------- [0:255 -1:255 30 0] [0x2 1 1 [1 0] [0 4]] (REQUEST)\\

s 21.500000000 \_0\_ RTR  --- 0 AODV 48 [0 0 0 0] ------- [0:255 -1:255 30 0] [0x2 1 4 [1 0] [0 12]] (REQUEST)\\

r 21.501260809 \_2\_ RTR  --- 0 AODV 48 [0 ffffffff 0 800] ------- [0:255 -1:255 30 0] [0x2 1 4 [1 0] [0 12]] (REQUEST)\\

The interpretation of the following trace format is as follows:\\

r 21.501260809 \_2\_ RTR  --- 0 AODV 48 [0 ffffffff 0 800] ------- [0:255 -1:255 30 0] [0x2 1 4 [1 0] [0 12]] (REQUEST)\\

Node ID 2, receives a packet type REQUEST (AODV protocol), at layer RTR (routing), at time 21.501260809. This packet have sequence number 0. \\

A generalized explanation of trace format ~\cite{trace} would be as follows : \\

\begin{tabular}{p{1.3cm} | p{3cm} | p{7cm}}
\hline Column Number & What Happened? & Values for instance... \\
\hline
\hline 1 & It shows the occured event & 's' SEND, 'r' RECEIVED, 'D' DROPPED \\ 
\hline 2 & Time at which the event occured? & 10.000000000 \\ 
\hline 3 & Node at which the event occured? & Node id like 0 \\
\hline 4 & Layer at which the event occured? & 'AGT' application layer, 'RTR' routing layer, 'LL' link layer, 'IFQ' Interface queue, 'MAC'  mac layer, 'PHY' physical layer \\
\hline 5 & show flags & --- \\
\hline 6 & shows the sequence number of packets &  0 \\
\hline 7 & shows the packet type & 'cbr' CBR packet, 'DSR' DSR packet, 'RTS' RTS packet generated by MAC layer, 'ARP' link layer ARP packet \\
\hline 8 & shows size of the packet & Packet size increases when a packet moves from an upper layer to a lower layer and decreases when a packet moves from a lower layer to an upper layer \\
\hline 9 & [....] & It shows information about packet duration, mac address of destination, the mac address of source,  and the mac type of the packet body. \\
\hline 10 & show flags & --- \\
\hline 11 & [....] & It shows information about source node ip : port number, destination node ip (-1 means broadcast) : port number, ip header ttl, and  ip of next hop (0 means node 0 or broadcast). \\
\hline  \\
\end{tabular}

\section{Main Implementation Files aodv.cc and aodv.h}
\label{sec:main}


\subsection{How to Enable Hello Packets}
\label{sec:hello}

By default HELLO packets are disabled in the aodv protocol. To enable broadcasting of Hello packets, comment the following two lines present in aodv.cc\\ \#ifndef AODV\_LINK\_LAYER\_DETECTION\\ \#endif LINK LAYER DETECTION 
and recompile ns2 by using the following commands on the terminal:\\
make clean\\
make \\
sudo make install\\

\subsection{Timers Used}
\label{sec:timers}
In ns2, timers are used to delay actions or can also be used for the repetition of a particular action like broadcasting of Hello packets after fixed time interval. Following are the timers that are used in AODV protocol implementation:

\begin{itemize}
\item Broadcast Timer: This timer is responsible for purging the ID's of Nodes and schedule after every BCAST\_ID\_SAVE.
\item Hello Timer: It is responsible for sending of Hello Packets with a delay value equal to interval, where \\
double interval = MinHelloInterval +
                 ((MaxHelloInterval - MinHelloInterval) *
Random::uniform()); 
\item Neighbor Timer: Purges all timed-out neighbor entries and schedule after every HELLO\_INTERVAL .
\item RouteCache Timer: This timer is responsible for purging the route from the routing table and schedule after every FREQUENCY.
\item Local Repair Timer: This timer is responsible for repairing the routes.
\end{itemize}

\vspace{20pt}

\subsection{Functions}
\label{sec:functions}

\subsubsection{General Functions}
\label{sec:genfun}
\begin{itemize}
\item void recv(Packet *p, Handler *): At the network layer, the Packet is first received at the recv() function, sended by the MAC layer in up direction. The recv() function will check the packet type. If the packet type is AODV type, it will decrease the TTL and call the recvAODV() function.\\
If the node itself generating the packet then add the IP header to handle broadcasting, otherwise check the routing loop, if routing loop is present then drop the packet, otherwise forward the packet.
\item int command(int, const char *const *): Every object created in NS-2 establishes an instance procedure, cmd\{\} as a hook to executing methods through the compiled shadow object. This procedure cmd{} invokes the method command() of the shadow object automatically, passes the arguments to cmd\{\} as an argument vector to the command() method \cite{IEEEhowto:nsmanual}.

\end{itemize}

\vspace{20pt}

\subsubsection{Functions for Routing Table Management}
\label{sec:funrouting}

\begin{itemize}
\item void rt\_resolve(Packet *p): This function first set the transmit failure callback and then forward the packet if the route is up else check if I am the source of the packet and then do a Route Request, else if the local repair is in progress then buffer the packet. \\

If this function founds that it has to forward a packet for someone else to which it does not have a route then drop the packet and send error upstream. Now after this, the route errors are broadcasted to the upstream neighbors.

\vspace{05pt}
\item void rt\_update(aodv\_rt\_entry *rt, u\_int32\_t seqnum,u\_int16\_t metric, nsaddr\_t nexthop,double expire\_time): This function is responsible for updating the route.
\vspace{05pt}
\item void rt\_down(aodv\_rt\_entry *rt): This function first confirms that the route should not be down more than once and after that down the route.
\vspace{05pt}
\item void local\_rt\_repair(aodv\_rt\_entry *rt, Packet *p): This function first buffer the packet and mark the route as under repair and send a RREQ packet by calling the sendRequest() function.
\vspace{05pt}
\item void rt\_ll\_failed(Packet *p): Basically this function is invoked whenever the link layer reports a route failure. This function drops the packet if link layer is not detected. Otherwise, if link layer is detected, drop the non-data packets and broadcast packets. If this function founds that the broken link is closer to the destination than source then It will try to attempt a local repair, else brings down the route.



\vspace{05pt}
\item void handle\_link\_failure(nsaddr\_t id): This function is responsible for handling the link failure. It first checks the DestCount, if It is equal to 0 then remove the lost neighbor. Otherwise, if DestCount $>$ 0 then send the error by calling sendError() function, else frees the packet up.
\vspace{05pt}

\item void rt\_purge(void): This function is responsible for purging the routing table entries from the routing table. For each route, this function will check whether the route has expired or not. If It founds that the valid route is expired, It will purge all the packets from send buffer and invalidate the route, by dropping the packets and tracing DROP\_RTR\_NO\_ROUTE "NRTE" in the trace file. If It founds that the valid route is not expired and there are packets in the sendbuffer waiting, It will forward them. Finally, if It founds that the route is down and if there is a packet for this destination waiting in the sendbuffer, It will call sendRequest() function.
\vspace{05pt}

\item void enque(aodv\_rt\_entry *rt, Packet *p): Use to enqueue the packet.
\vspace{05pt}
\item Packet* deque(aodv\_rt\_entry *rt): Use to dequeue the packet.

\end{itemize}

\vspace{20pt}

\subsubsection{Functions for Neighbors Management}
\label{sec:funneighbors}

\begin{itemize}
\item void nb\_insert(nsaddr\_t id): This function is used to insert the neighbor.
\item AODV\_Neighbor* nb\_lookup(nsaddr\_t id): This function is used to lookup the neighbor.
\item void nb\_delete(nsaddr\_t id): This function is used to delete the neighbor and It is called when a neighbor is no longer reachable.
\item void nb\_purge(void): This function purges all timed-out neighbor entries and It runs every \\
HELLO\_INTERVAL * 1.5 seconds.

\end{itemize}

\vspace{20pt}

\subsubsection{Functions for Broadcast ID Management}
\label{sec:funbroadcastid}

\begin{itemize}
\item void id\_insert(nsaddr\_t id, u\_int32\_t bid): This function is used to insert the broadcast ID of the node.
\item bool id\_lookup(nsaddr\_t id, u\_int32\_t bid): This function is used to lookup the broadcast ID.
\item void id\_purge(void): This function is used to purge the broadcast ID.
\end{itemize}

\vspace{20pt}

\subsubsection{Functions for Packet Transmission Management}
\label{sec:funtransmission}

\begin{itemize}
\item void forward(aodv\_rt\_entry *rt, Packet *p, double delay): This function is used to forward the packets.
\item void sendHello(void): This function is responsible for sending the Hello messages in a broadcast fashion.
\item void sendRequest(nsaddr\_t dst): This function is used to send Request messages.
\item void sendReply(nsaddr\_t ipdst, u\_int32\_t hop\_count,nsaddr\_t rpdst, u\_int32\_t rpseq,u\_int32\_t lifetime, double timestamp): This function is used to send Reply messages.
\item void sendError(Packet *p, bool jitter = true): This function is used to send Error messages.
\end{itemize}                                       

\vspace{20pt}

\subsubsection{Functions for Packet Reception Management}
\label{sec:funreception}

\begin{itemize}
\item AODV::recvAODV(Packet *p): This function classify the incoming AODV packets. If the incoming packet is of type RREQ, RREP, RERR, HELLO, It will call recvRequest(p), recvReply(p), recvError(p), and recvHello(p) functions respectively.

\item AODV::recvRequest(Packet *p): When a node receives a packet of type REQUEST, it calls this function.

\item AODV::recvReply(Packet *p): When a node receives a packet of type REPLY, it calls this function.

\item AODV::recvError(Packet *p): This function is called when a node receives an ERROR message.

\item AODV::recvHello(Packet *p): This function receives the HELLO packets and look into the neighbor list, if the node is not present in the neighbor list, It inserts the neighbor, otherwise if the neighbor is present in the neighbor list, set its expiry time to:\\
CURRENT\_TIME + (1.5 * ALLOWED\_HELLO\_LOSS * HELLO\_INTERVAL), where ALLOWED\_HELLO\_LOSS = 3 packets and HELLO\_INTERVAL = 1000 ms.

\end{itemize}

\vspace{20pt}

%
%
%
%


%
%
%
%

\vspace{20pt}
%
%
%

\bibliographystyle{IEEEtran}
\bibliography{aodvbib}

\pagebreak
\pagestyle{empty}

\section{Appendex : A Simple TCL Script to Run the AODV Protocol}
\label{sec:tcl}

\# wireless-aodv.tcl

\# A 3 nodes example for ad hoc simulation with AODV

\# Define options

set val(chan)    \hspace{20pt}       Channel/WirelessChannel;\# channel type

set val(prop)    \hspace{20pt}       Propagation/TwoRayGround;\# radio-propagation model

set val(netif)   \hspace{19pt}      Phy/WirelessPhy            ;\# network interface type

set val(mac)     \hspace{21pt}       Mac/802\_11                 ;\# MAC type

set val(ifq)     \hspace{25pt}       Queue/DropTail/PriQueue    ;\# interface queue type

set val(ll)      \hspace{30pt}       LL                         ;\# link layer type

set val(ant)     \hspace{24pt}       Antenna/OmniAntenna        ;\# antenna model

set val(ifqlen)  \hspace{13pt}       50                         ;\# max packet in ifq

set val(nn)      \hspace{26pt}       3                          ;\# number of mobilenodes

set val(rp)      \hspace{26pt}       AODV                       ;\# routing protocol

set val(x)       \hspace{28pt}       500   			           ;\# X dimension of topography

set val(y)       \hspace{28pt}       400   			           ;\# Y dimension of topography  

set val(stop)	\hspace{18pt}	150			                   ;\# time of simulation end

set ns	\hspace{46pt}	  [new Simulator]

set tracefd   \hspace{27pt}    [open simple.tr w]

set namtrace  \hspace{18pt}    [open simwrls.nam w]  
  
\$ns trace-all \$tracefd

\$ns namtrace-all-wireless \$namtrace \$val(x) \$val(y)

\# set up topography object

set topo       [new Topography]

\$topo load\_flatgrid \$val(x) \$val(y)

create-god \$val(nn)

\#  Create nn mobilenodes [\$val(nn)] and attach them to the channel. 

set chan\_1\_ [new \$val(chan)]

\# configure the nodes\\
        \$ns node-config -adhocRouting \$val(rp) \textbackslash \\
			 -llType \$val(ll) \textbackslash \\
			 -macType \$val(mac) \textbackslash \\
			 -channel \$chan\_1\_ \textbackslash \\
			 -ifqType \$val(ifq) \textbackslash \\
			 -ifqLen \$val(ifqlen) \textbackslash \\
			 -antType \$val(ant) \textbackslash \\
			 -propType \$val(prop) \textbackslash \\ 
			 -phyType \$val(netif) \textbackslash \\
			 -topoInstance \$topo \textbackslash \\
			 -agentTrace ON \textbackslash \\
			 -routerTrace ON \textbackslash \\
			 -macTrace OFF \textbackslash \\
			 -movementTrace ON \textbackslash \\

	for \{set i 0\} \{\$i \textless  \$val(nn) \} \{ incr i \} \{ \\
		set node\_(\$i) [\$ns node]	\\
	\} \\

\# Provide initial location of mobilenodes\\
\$node\_(0) set X\_ 5.0 \\
\$node\_(0) set Y\_ 5.0 \\
\$node\_(0) set Z\_ 0.0 \\
\$node\_(1) set X\_ 490.0 \\
\$node\_(1) set Y\_ 285.0 \\
\$node\_(1) set Z\_ 0.0 \\
\$node\_(2) set X\_ 150.0 \\
\$node\_(2) set Y\_ 240.0 \\
\$node\_(2) set Z\_ 0.0 \\

\# Generation of movements\\
\$ns at 10.0 ``\$node\_(0) setdest 250.0 250.0 3.0"\\
\$ns at 15.0 ``\$node\_(1) setdest 45.0 285.0 5.0"\\
\$ns at 110.0 ``\$node\_(0) setdest 480.0 300.0 5.0" \\

\# Set a TCP connection between node\_(0) and node\_(1) \\
set tcp [new Agent/TCP/Newreno] \\
\$tcp set class\_ 2 \\
set sink [new Agent/TCPSink] \\
\$ns attach-agent \$node\_(0) \$tcp \\
\$ns attach-agent \$node\_(1) \$sink \\
\$ns connect \$tcp \$sink \\
set ftp [new Application/FTP] \\
\$ftp attach-agent \$tcp \\
\$ns at 10.0 ``\$ftp start" \\

\# Define node initial position in nam\\
for \{set i 0\} \{\$i \textless  \$val(nn)\} \{ incr i \} \{\\
\# 30 defines the node size for nam\\
\$ns initial\_node\_pos \$node\_(\$i) 30\\
\}\\

\# Telling nodes when the simulation ends\\
for \{set i 0\} \{\$i \textless  \$val(nn) \} \{ incr i \} \{\\
    \$ns at \$val(stop) ``\$node\_(\$i) reset";\\
\}\\

\# ending nam and the simulation \\
\$ns at \$val(stop) ``\$ns nam-end-wireless \$val(stop)"\\
\$ns at \$val(stop) ``stop"\\
\$ns at 150.01``puts ``end simulation" ; \$ns halt"\\
proc stop \{\} \{\\
    global ns tracefd namtrace\\
    \$ns flush-trace\\
    close \$tracefd\\
    close \$namtrace\\
\}\\

\$ns run\\

\end{document}